\documentclass[final]{svjour2}
\usepackage{graphicx}
\usepackage{rotating}
\usepackage{amssymb}
\usepackage{amsmath}
\usepackage{mathptmx}
\usepackage[numbers]{natbib}
\makeatletter
\journalname{Journal of Low Temperature Physics}

\bibpunct{}{}{,}{s}{}{,}

\begin{document}

\newcommand{\hdblarrow}{H\makebox[0.9ex][l]{$\downdownarrows$}-}
\title{Cryogenic Silicon detectors with implanted contacts for the detection of visible photons using the Neganov-Trofimov-Luke Effect.}

\author{X. Defay\(^a\) \and E. Mondragon\(^a \) \and M. Willers\(^a\) \and A. Langenk\"{a}mper \(^a \)  \and  J.-C. Lanfranchi\(^a\) \and  A. M\"{u}nster\(^a \), A. Z\"{o}ller\(^a \) \and S. Wawoczny\(^a \) \and H. Steiger\(^a \) \and F. Hitzler\(^a \) \and C. Bruhn\(^a \) \and \linebreak S. Sch\"{o}nert\(^a\)\ \and W. Potzel\(^a \) \and M. Chapellier\(^b \)}

\institute{
\(^a \) : Excellence Cluster Universe and Physik Department, Technische Universit\"{a}t M\"{u}nchen, 85748 Garching, Germany.
\(^b \) : Centre de Sciences Nucleaire et de Sciences de la Matiere (CSNSM), IN2P3 Orsay, France.
}

\maketitle

\begin{abstract}

There is a common need in astroparticle experiments such as direct dark matter detection, double beta decay without emission of neutrinos $[$\(0 \nu \beta \beta \)$]$ and coherent neutrino nucleus scattering experiments for light detectors with a very low energy threshold. By employing the Neganov-Trofimov-Luke Effect, the thermal signal of particle interactions in a semiconductor absorber operated at cryogenic temperatures, can be amplified by drifting the photogenerated electrons and holes in an electric field. This technology is not used in current experiments, in particular because of a reduction of the signal amplitude with time which is due to trapping of the charges within the absorber. We present here the first results of a novel type of Neganov-Trofimov-Luke Effect light detector with an electric field configuration designed to improve the charge collection within the semiconductor.


\end{abstract}

\section{Introduction}

Bolometers represent a key technology in rare event experiments like direct dark matter search and \(0 \nu \beta \beta \) search. These experiments operate detectors at cryogenic temperatures to measure phonon-mediated signals. The threshold within these experiments is crucial. The Neganov-Trofimov-Luke Effect\cite{NeganovTrofimov,Luke,MauriceNTLE2000,MauriceFancy} (NTLE) can decrease the threshold of bolometric light detectors down to a few optical/UV photons. In  \(0 \nu \beta \beta \) experiments, this would allow to tag the Cherenkov light emitted by \(0 \nu \beta \beta \) events and reject completely the 
dominant $\alpha$ background, which is below the Cherenkov threshold emission. The rejection of the alpha background is necessary to extend the sensitivity of a bolometric experiment over the entire inverted hierarchy region of the neutrino mass pattern\cite{Artusa}. These light detectors can also enhance the sensitivity of dark matter searches\cite{CRESST2009} based on scintillating crystals by increasing the discrimination between signal and background events. In the context of direct dark matter searches, lowering the threshold represents a dramatic improvement of the sensitivity. This is particularly true for low mass WIMPs where the results of the experiments are currently characterised by contradictions between positive hints of a signal and total exclusion\cite{Davis,CRESSTLow}. 

Due to the NTLE, the threshold of low temperature light detectors based on semiconductor substrates can be improved significantly by drifting the photon-induced electron-hole pairs in an applied electric field, which results in additional heat. The following equation describes the total energy resulting in a thermal signal : 
\begin{equation}
	E_{Tot}=E_{0}\left(1+\frac{e \cdot V }{\epsilon}-\frac{\delta}{\epsilon}\right),
\end{equation}
where $E_{0}$ is the energy deposited by the particle, \(e \) the elementary charge, V the NTLE voltage,  $\delta$ the band gap (1.17 eV for Silicon\cite{BandGap}) and \(\epsilon \) the average energy required to create an electron-hole pair (which is equal to the photon energy for scintillation photons). The amplitude of the thermal signal is proportional to the voltage applied across the semiconductor while the noise is typically independent of it. Therefore, the signal-to-noise ratio can be increased using the NTLE.

\section{A novel Neganov-Trofimov-Luke effect light detector}

NTLE detectors consist of a semiconductor absorber (Silicon or Germanium) with two electrodes to apply a voltage, and a thermal sensor installed on the semiconductor. In the present work a Transition Edge Sensor (TES) is used as a heat sensor. For most of the NTLE detectors, metallic electrodes are deposited on the surface of a semiconductor absorber\cite{Stark2005} which implies that the charges drift along the free surfaces of the semiconductor (i.e., where no metal is deposited). However, the trapping cross-sections on the free surfaces are high\cite{Kingston}, this results in a bad charge collection and the amplitude of the thermal signal decreases in time\cite{Isaila2012}.

\begin{figure}
\begin{center}
\includegraphics[width=0.6\linewidth,totalheight=0.2\textheight]{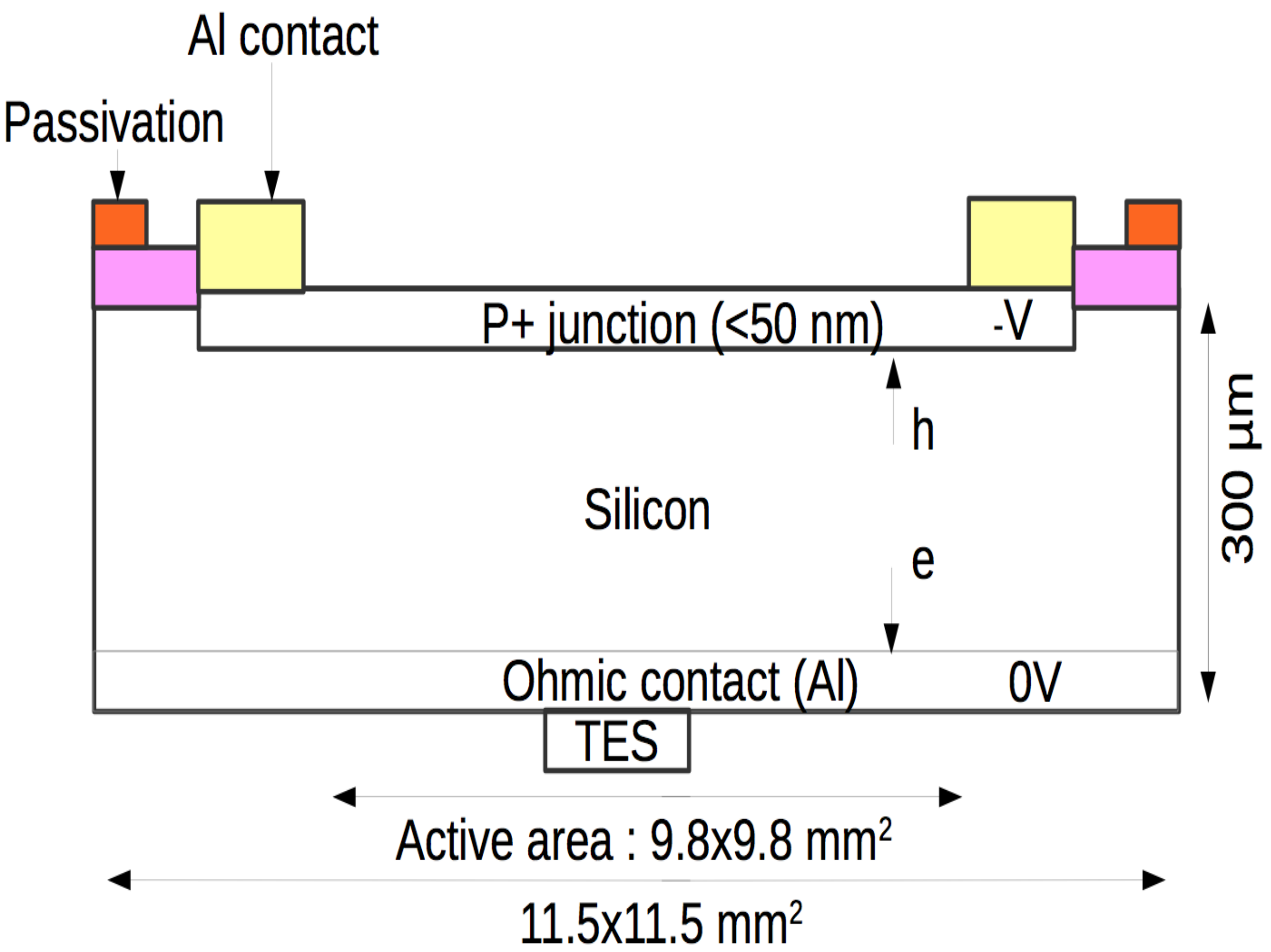}
\end{center}
\caption{Schematics of the NTLE detector. The TES, which is a Tungsten film deposited on a silicon carrier, is glued on the Aluminium contact of the NTLE absorber.}
\label{fig1}
\end{figure}

In the present work, we use a novel detector design presenting no free surfaces. This aims at the suppression of the degradation using a planar geometry shown in Fig.~\ref{fig1}. The entrance window consists of a very shallow Boron implanted contact featuring a high transmittance for 430 nm photons. The bulk of the NTLE detector is made of high-purity intrinsic silicon and the back electrode consists of an ohmic contact with an Aluminium layer. 

This approach is more favourable in terms of trapping for two reasons : 
\begin{itemize}
\item All surfaces are covered with electrodes avoiding the trapping on free surfaces.
\item The electric field within the detector is very high : no leak current was detected even for an electric field as high as 6500 V$\cdot$cm\(^{-1}\). This tremendous electric field favours a good collection of the carriers.
\end{itemize}

\section{Thermal gain and signal-to-noise improvement}

\begin{figure}
\begin{center}
\includegraphics[width=0.49\linewidth,keepaspectratio]{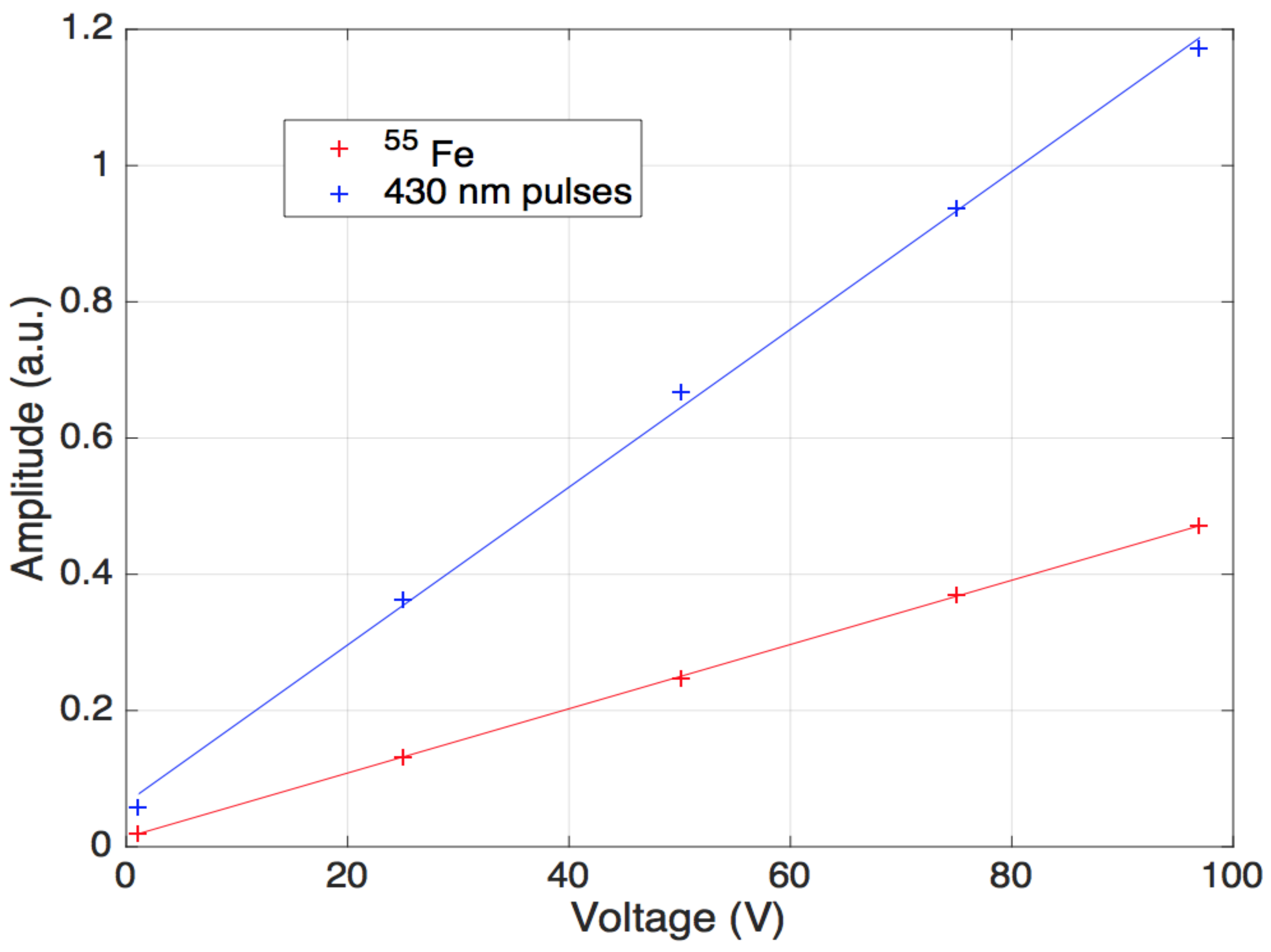}
\includegraphics[width=0.49\linewidth,keepaspectratio]{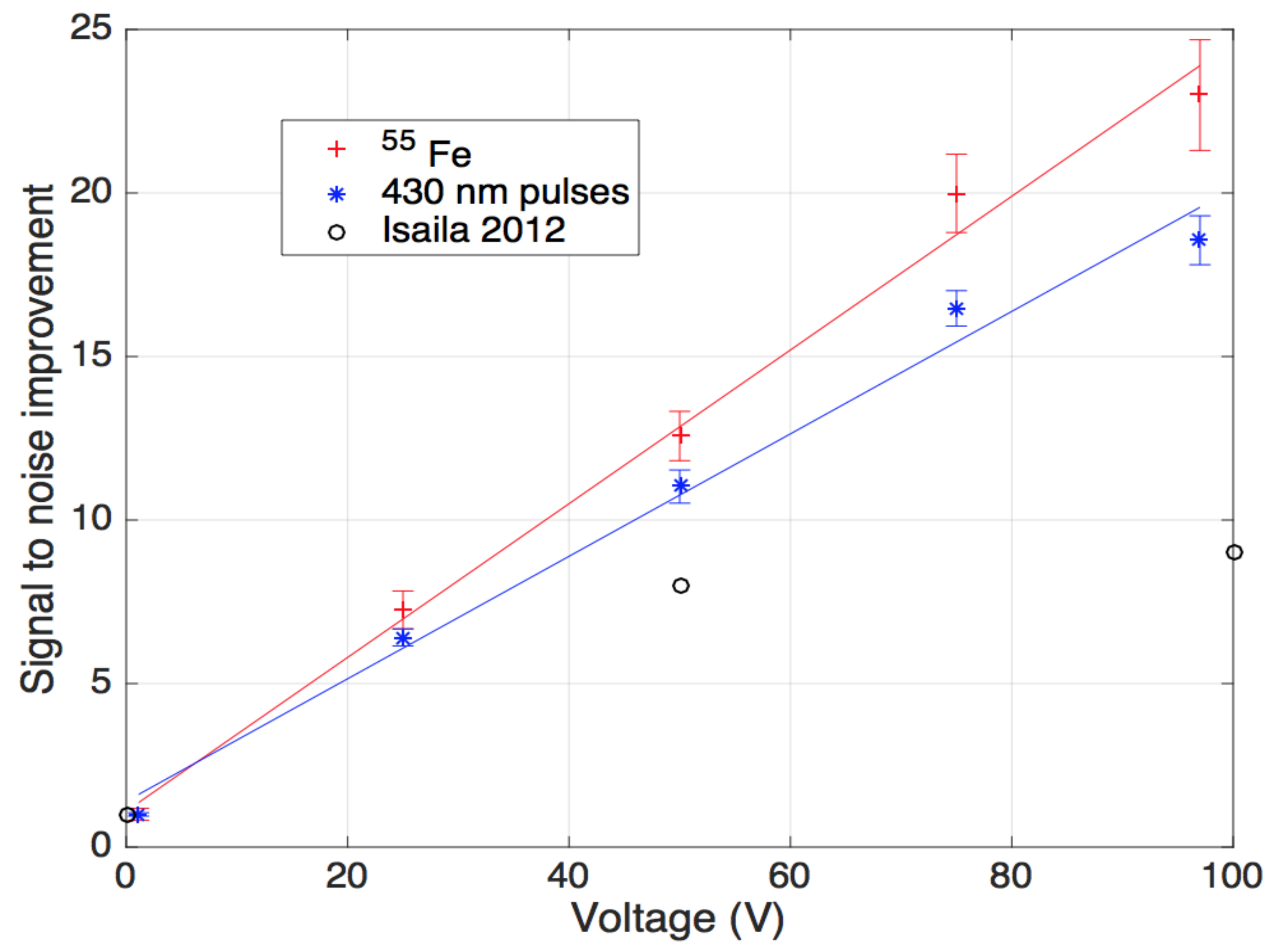}
\end{center}
\caption{{\it Left :} Amplitude of the signal as a function of the NTLE voltage for \(^{55} \)Fe and 430 nm  photon flashes (the error bars are smaller than the symbols used). {\it Right :} Signal-to-noise improvement versus the NTLE voltage (the error bars correspond to $1\sigma$). The result from the previous detector technology for 430 nm photons is symbolised by empty circles\cite{Isaila2012}. }
\label{fig2}
\end{figure}

The detector was irradiated with 430 nm  photons (matching the CaWO\(_{4}\) scintillation properties) together with X-rays at 5.9 and 6.4 keV emitted by an \(^{55} \)Fe (Mn X-ray) calibration source. The optical photons are generated by an LED operated outside the cryostat and guided onto the light detector using an optical fibre which illuminates the whole detector surface. The response of the phonon signals read on the TES was recorded while varying the NTLE voltage (Fig.~\ref{fig2} {\it Left}). For both the X-rays and 430 nm photons, the amplitudes of the thermal pulses increases linearly with the applied NTLE voltage. 

During all measurements, the amplitude of the noise remained independent of the NTLE voltage applied. Therefore, the signal-to-noise ratio (amplitude of the thermal signal divided by the $1\sigma$ width of the baseline noise) increases proportionally to the NTLE voltage (Fig.~\ref{fig2} {\it Right}). This represents a significant improvement compared to the previous detector technology \cite{Isaila2012}. 
The $1\sigma$ width of the baseline noise at 97V calibrated with the \(^{55} \)Fe calibration line corresponds to 5.33 $\pm$ 0.41 eV.

\section{Degradation study}

The signal-to-noise improvement is very high and our main concern focuses on the collection of the carriers. After the creation of the electron-hole pairs, the electrons and holes drift in opposite directions. Along the way, if these charges are trapped, the NTLE signal will not be fully produced. The trapped charges end-up building a counter electric field which tends to suppress the initial electric field within the detector. This so-called degradation effect is a major concern since it implies that some particle events are not well measured. The behaviour of the detector evolves in time and depends on the  position of the interaction within the detector. A regeneration procedure must be applied periodically to recover the initial charge collection\cite{Isaila2012}. This is the main reason why NTLE detectors are not used in current experiments. Dark matter and \(0 \nu \beta \beta \) experiments are looking for very rare signals of a few photons and require a very high degree of reliability of the detectors. 

Therefore, a dedicated experiment was performed to study the degradation of our detector. The amplitude of the counter electric field due to the degradation depends on the amount of charges created within the detector. In order to evaluate this degradation, we decided to expose the detector to a very high flux of photons and to study the evolution of the amplitude of the phonon signal. 

\begin{figure}
\begin{center}
\includegraphics[width=0.64\linewidth,keepaspectratio]{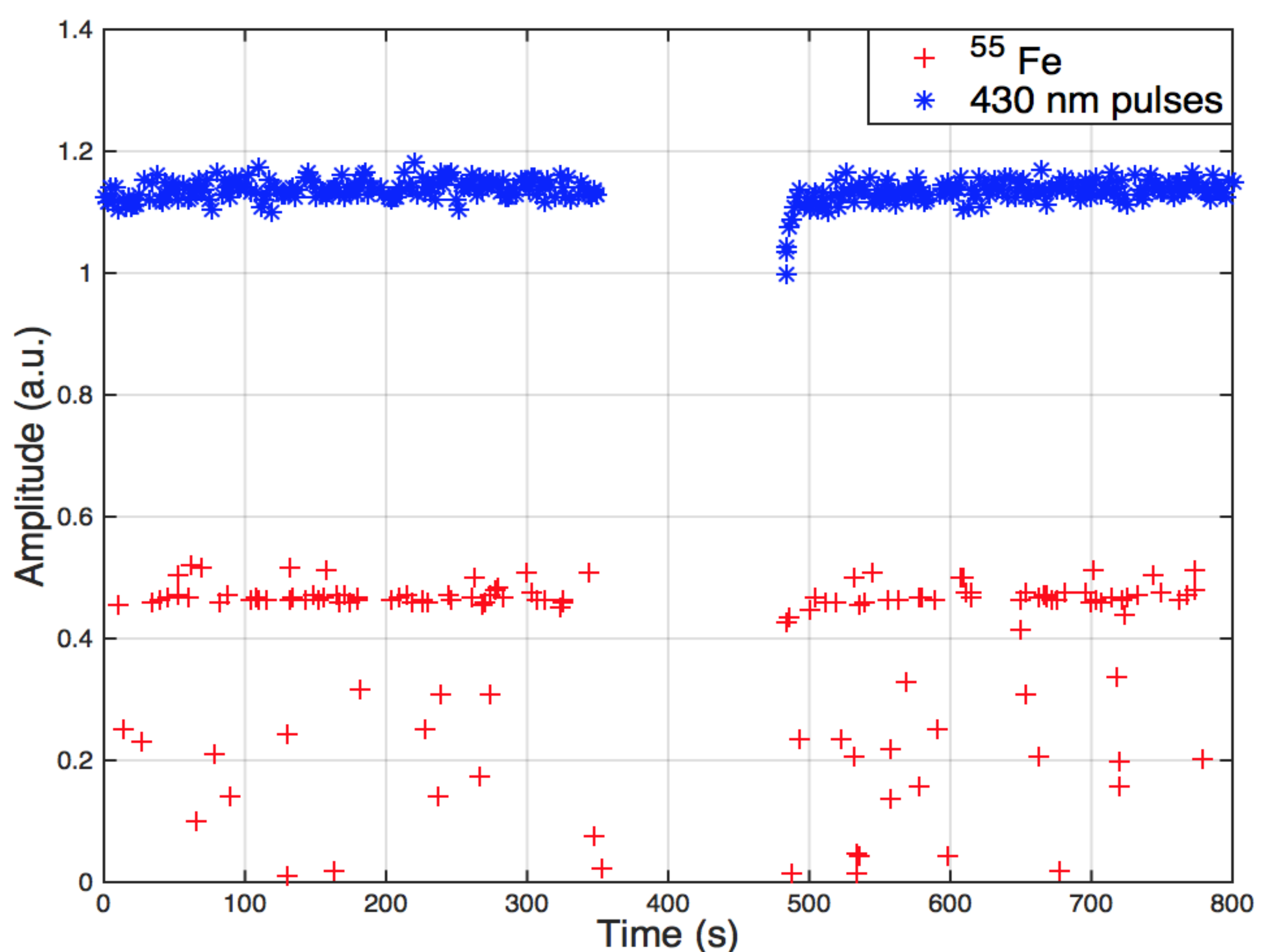}
\end{center}
\caption{Evolution of the amplitude of the heat signals during a degradation experiment (for a detailed description refer to the main text). During this experiment the NTLE voltage is kept at 97V all the time. }
\label{fig3}
\end{figure}

This experiment consists of 3 phases (Fig.~\ref{fig3}) : 
\begin{enumerate}
\item The detector is first calibrated with an \(^{55} \)Fe source and 430 nm photon flashes at a rate of 0.5 Hz and a NTLE voltage of 97 V applied. 
\item The flux of 430 nm photons is multiplied by a factor of 10000 (which corresponds to a power of 100 MeV$\cdot$s\( ^{-1} \)) in order to create a large amount of charges. During this phase, no signal can be measured from the TES because the signals consist of pile-up events only. The NTLE voltage is kept at 97 V. The large amount of charges created constitutes extreme degradation conditions. 
\item During the final phase of the experiment, the flux of photon pulses is decreased to its initial value (0.5 Hz). After a few seconds the signals recover to their initial amplitude which means that no degradation was observed. 
\end{enumerate}
This result is very important for the application of NTLE detectors in rare event experiments since it implies that such a detector, in contrast with previous NTLE detectors, does not need any charge regeneration procedure and presents a very reliable charge collection.

\begin{figure}
\begin{center}
\includegraphics[width=0.64\linewidth,keepaspectratio]{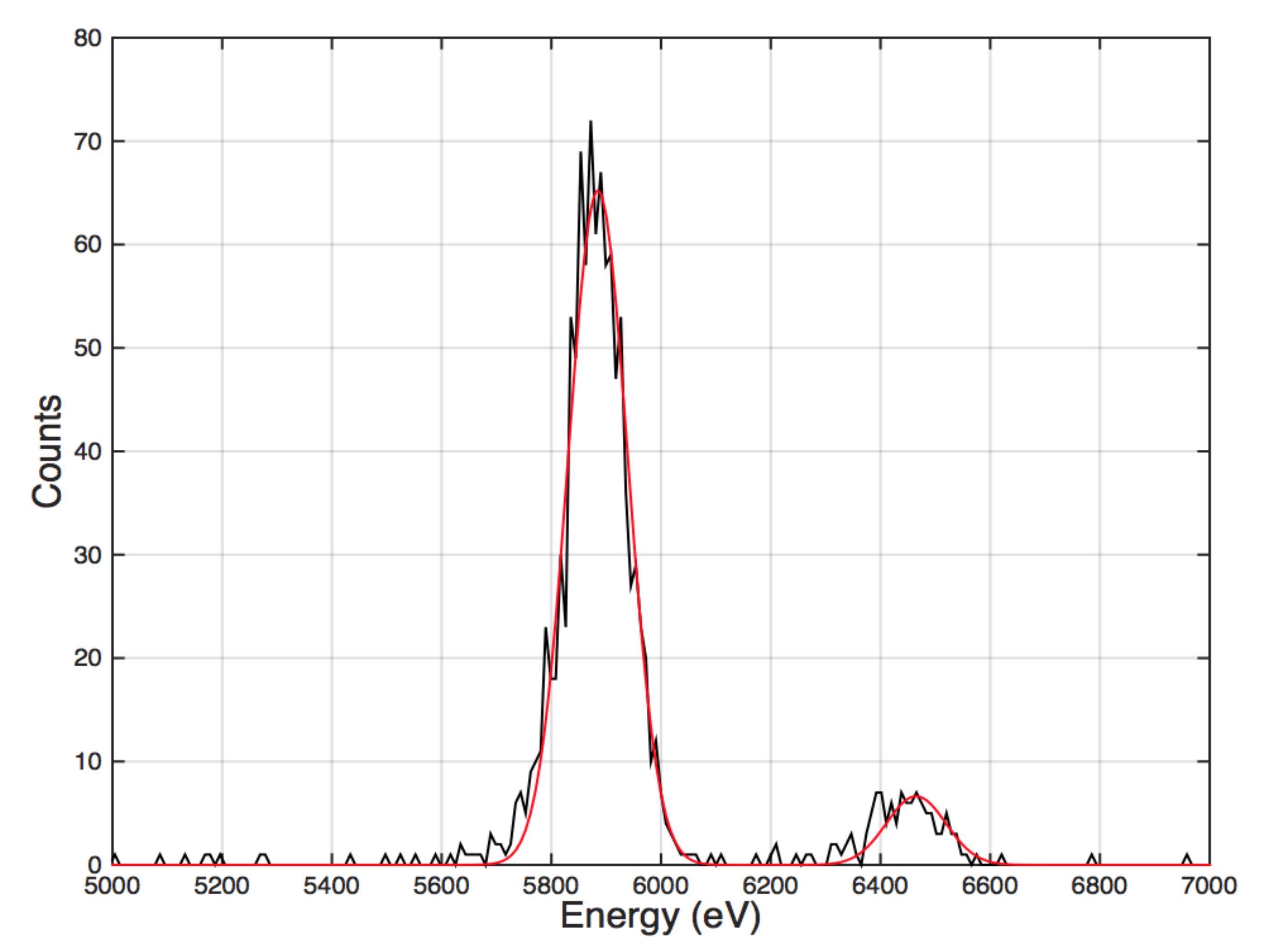}
\end{center}
\caption{Histogram of the X-ray events from the \(^{55} \)Fe (Mn X-rays) source. Experimental data in black and gaussian fit in red.}
\label{fig4}
\end{figure}

More information on the collection of the carriers can be inferred from the energy width of the X-ray events. The theoretical resolution of the X-ray events can be evaluated as follow : 
\begin{equation}
	\Delta E = 2.35 \cdot \sqrt{E \cdot \epsilon \cdot F} ,
\end{equation}
where $E$ is the energy of the event, $\epsilon$ the energy required to create an electron-hole pair and $F$ the Fano factor. The theoretical resolution is 120 eV using  $\epsilon$=3.8 eV\cite{knoll}  and $F$=0.1161\cite{Lowe}. The experimental resolution of the detector at 5.9 keV and 6.4 keV is (127.4 $\pm$ 3.8) eV (Fig.~\ref{fig4}). The fact that this number is very close to the theoretical one implies that the collection of the events is homogeneous on the whole surface of the detector.

\section{Conclusion}

The novel NTLE detector studied in the present work features a high thermal gain which allows us to drastically improve the signal-to-noise ratio for both X-rays and 430 nm photons. Furthermore, no degradation was observed even after injecting a large amount of charges which demonstrates that this detector is a promising solution for cryogenic experiments requiring the detection of photons at very low energy.

\begin{acknowledgements}

This research was supported by the DFG cluster of excellence "Origin and Structure of the Universe" (www.universe-cluster.de).

\end{acknowledgements}

\pagebreak


\begin{thebibliography}{99}

\bibitem{NeganovTrofimov}
B. Neganov and V. Trofimov, {\it Otkryt. Izobret.} \textbf{146}, 215, (1985).

\bibitem{Luke}
P.N. Luke, Voltage-assisted calorimetric ionization detector, {\it J. Appl. Phys.} \textbf{64}, 6858, (1988).

\bibitem{MauriceNTLE2000}
M. P. Chapellier, G. Chardin, L. Miramonti, X. F. Navick, Physical interpretation of the Neganov-Trofimov-Luke and related effects, {\it Physica B} \textbf{284-288}, 2135-2136, (2000).

\bibitem{MauriceFancy}
M. P. Chapellier, Fancy Ideas on Neganov-Trofimov-Luke Effect (NTL) or: Is There a Limit to the NTL Amplification?, {\it J Low temp Phys} \textbf{237-242}, 178, (2015).

\bibitem{Artusa}
CUORE Collaboration, Initial performance of the CUORE-0 experiment, {\it Eur.Phys.J.} \textbf{C74}, 8, 2956, (2014).

\bibitem{CRESST2009}
G. Angloher et al., Commissioning run of the CRESST-II dark matter search, {\it Astropart. Phys.} \textbf{31}, 270, (2009).

\bibitem{Davis}
J. H. Davis, The Past and Future of Light Dark Matter Direct Detection, {\it Int. J. Mod. Phys. A} \textbf{30}, 15, (2015).

\bibitem{CRESSTLow}
The CRESST collaboration, Results on light dark matter particles with a low-threshold CRESST-II detector, {\it Eur. Phys. J. C} \textbf{75}, (2015).

\bibitem{BandGap}
K. P. O'Donnell and X. Chen, Temperature dependence of the semiconductor band gaps, , {\it Appl. Phys. Lett.} \textbf{58}, (1991).

\bibitem{Stark2005}
M. Stark et al., Application of the Neganov-Trofimov-Luke effect to low-threshold light detectors, {\it Nucl. Instr. Meth. Phys. Res. A} \textbf{545}, 738, (2005).

\bibitem{Kingston}
R.H. Kingston, {\it J. Appl. Phys.} \textbf{27}, 101, (1956).

\bibitem{Isaila2012}
C. Isaila et al., Low-temperature light detectors: Neganov-Trofimov-Luke amplification and calibration, {\it Phys.Lett. B} \textbf{716}, 160, (2012).

\bibitem{knoll}
G.F. Knoll, Radiation Detection and Measurement, 3rd ed., John Wiley \& Sons, New York (2000).


\bibitem{Lowe}
B.G. Lowe, Measurements of Fano factors in Silicon and Germanium in the low-energy X-ray region {\it Nucl. Instr. Meth. Phys. Res. A} \textbf{339}, 354, (1997).


\end{thebibliography}
\end{document}